\documentclass[aps,prb,twocolumn,amsmath,amssymb]{revtex4}
\usepackage{graphicx}
\usepackage{bm}
\usepackage{multirow}
\usepackage{times}

\begin{document}
\title{Nano-pattern induced ferromagnetism in strongly correlated electrons}
\author{Arnaud Ralko$^1$ and Georges Bouzerar$^{1,2}$}
\affiliation{
$^1$Institut N\'eel, UPR2940, CNRS et Universit\'e de Grenoble, Grenoble, FR-38042
France
\\ $^{2}$Jacobs University Bremen, Campus Ring 1, D-28759 Bremen, Germany} 
\date{\today}

\begin{abstract}
Band ferromagnetism in strongly correlated electron systems is one of the most
challenging issue in today's condensed-matter physics.
In this theoretical work, we study  the competition between kinetic term,
Coulomb repulsion, and on-site correlated disorder for various lattice
geometries.  Unconventional and complex ferromagnetic phase diagrams are
obtained: wide region of stability, cascade of transitions, re-entrance, high
sensitivity to the carrier concentration and strongly inhomogeneous ground
states for relatively weak on-site potential. 
The direct and systematic comparison with Exact Diagonalization shows that the
Unrestricted Hartree-Fock method is unexpectedly accurate for such systems, 
which allows large size cluster calculations. A match of the order of 99.9\%  for
weak and intermediate couplings is found, slightly reduced to about 95\% in the
large repulsion regime. 
Nano-patterned lattices  appear to be particularly promising candidates that
could, with the tremendous progress in growing and self-organized techniques,
be synthesized in a near future.
\end{abstract}
\maketitle

The possibility of ferromagnetic ground state (GS) in strongly correlated
itinerant systems is a highly non trivial question and is still one of the most
debated major issue.
The minimal Hamiltonian, introduced fifty years ago by Gutzwiller, Hubbard and
Kanamori in order to tackle such a subtle physics, is known as the Hubbard
model \cite{hubbard-model}.
More precisely, one of the main issue was the better understanding of the
appearance of ferromagnetism in transition metals.
The Hubbard model was introduced fifty years ago by Gutzwiller, Hubbard and
Kanamori in order to get a better understanding of the appearance of
ferromagnetism in transition metals \cite{hubbard-model}.
A first rigorous example of ferromagnetism was provided by Nagaoka and Thouless
\cite{nagaoka,thouless} a couple of years later;  a certain class of Hubbard
models exhibit saturated ferromagnetic GS for infinite Coulomb
repulsion at half filling and under the condition of having exactly one single
hole in the system.
The question of whether Nagaoka ferromagnetism survives at finite repulsion and
at larger hole density has not been answered so far.
More than twenty years later, Lieb established a theorem \cite{lieb} stating
that a system of half-filled bipartite lattices,  with different numbers of
sublattice sites, possesses a unique ferromagnetic GS (in fact
ferrimagnetic).
In the same period, another important rigorous result was obtained by Mielke
\cite{mielke} and Tasaki \cite{tasaki}. They formulated a theorem stating
that systems with nearly flat and partially filled band have a global stable
ferromagnetic GS.
For example, a very beautiful striking example of this kind of ferromagnetism
is achieved on the Kagome lattice (3 atoms per unit cell).
It is worth noticing that Lieb ferrimagnetism appears to be a particular case
of Mielke-Tasaki's magnetism at half filling.
One remarkable feature of this theorem is that even an infinitesimal coulombian
repulsion $U$ leads to a fully spin polarized GS while the magnetism is induced
by the electronic correlations.
More recently, several numerical studies have shown that 1D lattices based on
connected triangles are favourable for ferromagnetism
\cite{penc,watanabe,derzhko2010}.
This emphasizes the important role of the frustration in the statibilization of
ferromagnetism.
In the same spirit, the possibility of ferromagnetism in quantum dot
super-lattices \cite{tamura2002} and organic polymers consisting of chains of
five-membered rings \cite{gulacsi2010} have been investigated leading to
possible experimental realizations as the polyaminotriazole \cite{arita2002} or
polymethylaminotriazole \cite{suwa2003}. 

\begin{figure}[h] 
\includegraphics[width=0.45\textwidth,clip]{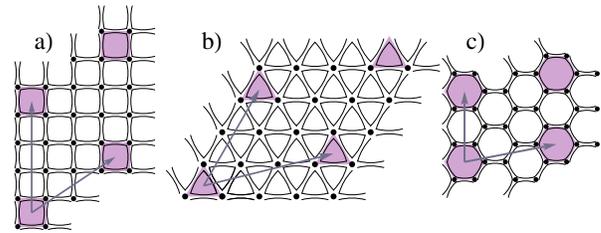}
\caption{(Color online). 
The three geometries of the nano-patterned lattices considered in this work:
(a) square, (b) triangular and (c) graphene.
The defect plaquette (shadded polygons) sublattice is defined by the elementary
vectors (arrows).
\label{Fig01}
}
\end{figure}

Another exotic type of ferromagnetism has been recently reported in oxide thin
films, {\it e.g.} HfO$_2$ \cite{coey} and CaO \cite{elfimov}, also known as
d$^0$ ferromagnetism. In this case, the presence of non magnetic defects
(either vacancies, cationic and/or anionic impurities) induces finite magnetic
moment in their vicinity and eventually long range ferromagnetic order. 
These unexpected findings have been followed by numerous debates
\cite{coey,elfimov,venkatesan,pemmaraju,zunger}.
A simple interpretation of this d$^0$ ferromagnetism has been proposed within a
single orbital Hubbard model with short range correlated disorder on the oxygen
orbitals by using an unrestricted Hartree-Fock method \cite{bouzerar2006}.
This study also led to interesting predictions; the compound
A$_{1-x}$B$_{x}$O$_2$, A  being either Zr, Ti, Hf, etc, doped by K or Na, could
provide high critical temperature ferromagnetic systems, as confirmed later on
by ab-initio based calculations \cite{maca2008}. 
Finally, this scenario has also explained the possibility of ferromagnetic
phases in graphene and irradiated graphite \cite{rossier2007,yazyev2008}.
The recent developments in nanotechnology growth and self-organized techniques
are opening new paths towards the realization of quasi 1D and 2D
super-lattices, nano-cluster arrays \cite{li} or decorated lattices
\cite{nano123}. Hence, ferromagnetism in strongly correlated electron systems
is still an uncharted terrain worth to be explored.

In the above mentioned numerical studies, the crucial ingredient is the
correlated nature of the disorder. For example, in K doped ZrO$_2$, the
substitution of the cations Zr$^{4+}$ by $K^+$ provides three holes per
impurity and induces short range correlated onsite disorder on the neighboring
oxygen orbitals. 
Thus, the main motivation of this study is to analyse the effects of
short-range correlated disorder in two dimensional interacting electron systems.
More precisely, we investigate the possibility of ferromagnetism on various
nano-patterned lattices.
In order to draw general conclusions, we have considered three different
geometries: square, triangular and honeycomb (graphene) lattices, as
depicted in Fig.~\ref{Fig01}.
Note that the origin of such defect plaquettes in real material could be the
substitution, presence of add-atoms or vacancies, or other structural defects.

The minimal model for that purpose is the one-orbital Hubbard Hamiltonian :
\begin{equation}
\label{Eq:ham} H = -t \sum_{<i,j>}  c_i^\dagger c_j +  U \sum_i n_{i,\uparrow} n_{i,\downarrow} + V \sum_{i \in
\textrm{def.},\sigma}   n_{i,\sigma}
\end{equation}
where $c_i^\dagger = (c_{i,\uparrow}^{\dagger}, c_{i,\downarrow}^{\dagger})$ is
a spinor of fermion {(electron)} creation operators at site $i$ and
$n_{i,\sigma} = c_{i,\sigma}^\dagger c_{i,\sigma} $ the on-site density operator. Note
that the hopping term $t$ and the on-site Coulomb repulsion $U$ have the same
magnitude on all bonds and sites respectively; the on-site spin independent
scattering term $V$ is defined only on sites belonging to certain defect
plaquettes of the nano-patterned lattice as depicted in Fig.~\ref{Fig01}.
In what follows, $U$ and $V$ will be expressed in units of $t$.
Defect plaquette sublattices on the three geometries considered in this work
are also displayed. These super-lattices are defined by unit-cell vectors
$({\bf t}_1,{\bf t}_2)$, both expressed in the units of the underlying lattice
vectors. For example, in Fig.~\ref{Fig01}-b (triangular lattice), these vectors
are respectively ${\bf t}_1 =(3,1)$ and ${\bf t}_2 =(0,3)$.

To handle such kind of models, various numerical methods have been employed
during years, all with advantages and limitations.
On one side, exact diagonalization (ED) allows an exact treatment but is
restricted to small clusters only due to an exponential growth of the Hilbert
space.
On an other side, within Quantum Monte-Carlo (QMC) based  methods, larger sizes
are reachable, but the sign problem can occur.
Finally, density matrix renormalization group (DMRG) can circumvent these
limitations, but remains essentially suitable for 1D systems.

From the mean field approaches, it is well known that the standard (restricted)
Hartree Fock approximation, which leads to the Stoner criterion ($U \rho_{F}
\ge 1$, with $\rho_F$ the single particle density of states at the Fermi
level), overestimates the tendency toward ferromagnetism in most systems
 \cite{hirsch}. 
It is worth noting that in Nagaoka, Lieb and  Mielke-Tasaki ferromagnetism,
this criterion is always fulfilled  since $\rho_{F}=\infty$.
However, the unrestricted Hartree Fock approach (UHF) has been successfully
employed in the  description of some exact GS properties in the
presence of disorder.
In this approximation, the interaction  term is decoupled as
\begin{eqnarray}
n_{i,\uparrow} n_{i,\downarrow}  \rightarrow  \langle n_{i,\uparrow}\rangle
n_{i,\downarrow} +  n_{i,\uparrow} \langle n_{i,\downarrow} \rangle -  \langle
n_{i,\uparrow} \rangle \langle n_{i,\downarrow} \rangle
\end{eqnarray} and the set $ \langle n_{i,\sigma}\rangle $, for $i \in [1, N]$ with $N$ the
number of sites, is calculated self-consistently. 
It is worth noticing that such an approach is, due to the self consistency,
non-linear and non-perturbative in essence;  for a 2D square lattice at half
filling for example, one gets a gap of the form $\Delta_U \sim e^{- 2 \pi
\sqrt{t/U}}$, as obtained in \cite{hirsch}.
As a first example, a fairly good agreement between this method and ED has been
obtained in the case of persistent currents of disordered mesoscopic rings
\cite{gb-pc}.
The comparison with QMC also supported the reliability of the UHF for the study
of disordered superconducting systems \cite{trivedi}.
In these works, the localized and/or inhomogeneous nature of the one particle
wave functions is at the origin of the fact that the Slater determinant becomes
a good estimate of the exact many-body GS.
It is worth noticing that the UHF usually provides a reasonable order of
magnitude of the physical observables and a rather good qualitative description
of phase diagrams.
Surprisingly enough, in spite of the expected richness of the physics, the
validity of the UHF approach has not been provided so far for
multi-band/orbital systems.
We also expect in the case of super-lattices (several atoms per unit cells),
giving rise to complex multi-band structure, the presence of relatively flat
band regions which should favor the appearance of ferromagnetism.
Our goal is now to investigate such a possibility in nano-patterned lattices.
Simultaneously, we will establish the accuracy of the UHF approach by direct
systematic comparison with ED calculations up to the largest accessible system
({\it e.g.} $8 \times 8$ sites). In addition, it should be stressed that the
agreement between ED and UHF is not only achieved for weak and intermediate
Hubbard repulsions, but also for large $U$. This will be shown in what follows.

\begin{figure}[h] 
\vspace{0.015\textwidth}
\includegraphics[width=0.50\textwidth,clip]{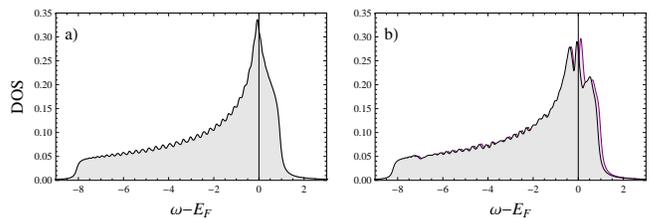}
\caption{
(Color online). Spin resolved density of states on a  $72 \times 72$
site cluster for respectively $(U,V)= (0,0)$ (a) and $(U,V) = (1.5,1)$ (b).
The carrier concentration is $n=1.5625$.  The vectors characterizing the
pattern are ${\bf t}_1 = (2,2)$ and ${\bf t}_2 = (-2,2)$ (see
Fig.~\ref{Fig01}).
\label{Fig02}
}
\end{figure}

 Let us first start with the effects of nano-patterns on the one-particle
 density of states (DOS). For that purpose, we performed the UFH calculation on
 a $72 \times 72$ triangular site cluster, at a carrier concentration close to
 the Van Hove singularity of the clean case, namely $n = 3/2 + \epsilon$, with
 $\epsilon = 0.0625$ (see Fig.\ref{Fig02}). The $(U,V)$ parameters are set to
 $(1.5,1)$. As can be seen, for finite $U$ and $V$, the ground state is
 ferromagnetic. Beside this band splitting, one clearly sees in the vicinity at
 the Fermi level, that the Van Hove singularity is now replaced by a
 double-peak structure. This feature survives in the thermodynamic limit.
Unfortunately, the DOS is not the appropriate quantity to provide any
information on the nature of the many body ground state.
Furthermore, as pointed out before, the standard HF is known to overestimate
the tendency to ferromagnetism.
Thus, one of the crucial questions is whether or not the true GS is really
ferromagnetic? In other words, is the UFH approach reliable for such systems?
 %
 %
 As it will be seen in what follows, even for such relatively small parameters,
 the GS turns out to be very inhomogeneous with strong fluctuations in local charge
 and spin densities, not reflected by the DOS.

\begin{figure}[h] 
\vspace{0.015\textwidth}
\includegraphics[width=0.50\textwidth,clip]{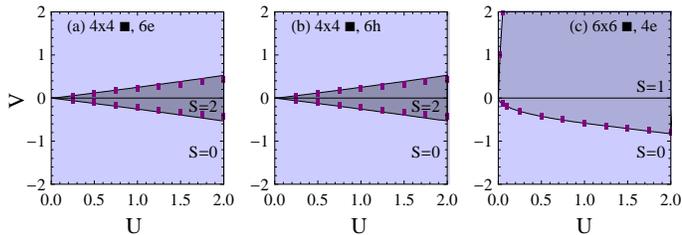}
\caption{
(Color online). Ground state phase diagrams for the square lattice obtained by
ED (symbols) and UHF (continuous lines) methods.
The number of sites $N = l \times l$ with $l$, the
carrier (electron/hole) number and the total spin $S$ are indicated.
In each case, we considered 2 defect plaquettes separated by a vector (2,2) in
units of their respective lattice vectors (see Fig.~\ref{Fig01}).
\label{Fig03}
}
\end{figure}

\begin{figure}[h] 
\vspace{0.015\textwidth}
\includegraphics[width=0.50\textwidth,clip]{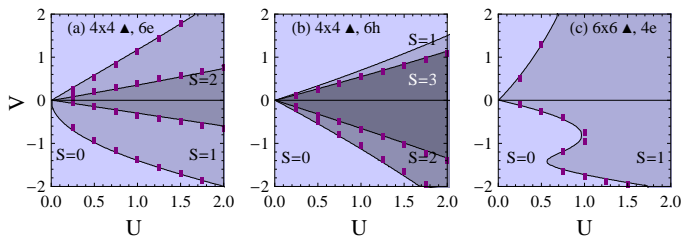}
\caption{
(Color online). Ground state phase diagrams for the triangular lattice
obtained by ED (symbols) and UHF (continuous lines) methods (see Fig.\ref{Fig03}
for notations).
\label{Fig04}
}
\end{figure}

In order to reveal these drastic effects, we propose to proceed with a
systematic comparison between ED and UHF calculations.
We start with the $(U,V)$ phase diagrams for both electron/hole doped
square and triangular lattices.
The results are plotted in Fig.~\ref{Fig03}
and Fig.~\ref{Fig04} respectively.
In all cases, the systems contain two defect plaquettes.
Phase diagrams on the square lattice for both electrons and holes at filling $n
= \frac{3}{8}$ in Fig.~\ref{Fig03}(a)-(b) look identical, as
expected for systems presenting the electron/hole symmetry (electron
$\leftrightarrow$ hole leads to $V \leftrightarrow -V$).
First of all, one immediately sees a remarkable agreement between the ED
critical lines and the UHF ones.
We find an extended large spin GS $S=2$ sector (the largest possible
spin being $S=3$ in this case), separated from a singlet region ($S=0$) without
any intermediate triplet phase.
The critical line of this phase transition is well approximated by $V_c \simeq
U/4$, hence showing that for a given $V$, a reasonable Hubbard repulsion is
enough to stabilize a ferromagnetic phase.

We now consider, for the same carrier density and same cluster size, the case
of the triangular lattice depicted in Fig.~\ref{Fig04}(a)-(b).
The phase diagrams appear to be more complex and richer.
In contrast to the square lattice, and as a consequence of the particle/hole
symmetry breaking, they are now completely different. 
In the case of electron doped, for a given $U$, a cascade from $S=2$ sector to
the singlet one through an intermediate  $S=1$ region is obtained.
However, the hole doped case behaves differently.
We indeed observe a wide fully polarized $S=3$ region, and for a fixed value of
$U$, three successive phase transitions at $V_{c,1}$, $V_{c,2}$ and $V_{c,3}$.
These critical values correspond respectively to  magnetic transitions $S_i \to
S_j$ of $0 \to 2 \to 3 \to 0$ when tuning $V$ from negative to positive values.
Remark that a fourth additional transition between $3 \to 1$ sectors at $V>0$
is obtained within the UHF. This implies the presence of a $S=1$ narrow region
which we interpret as a consequence of the quasi-degeneracy of the $S=0$ and
$S=1$ GS energies, as verified in the ED spectrum.
However, the agreement between the two methods remains excellent in both cases
of electron and hole doped systems and surpasses our initial expectations.
Let us now consider larger cluster sizes.  Results for the $6 \times 6$ site
cluster filled with 4 electrons ($n = 1/9$) are depicted in
Fig.~\ref{Fig03}(c) and Fig.~\ref{Fig04}(c) for respectively square and
triangular lattice. 
%
%
In contrast to panels (a) and (b), the phase diagram plotted in (c) is
now completely asymmetric. It is not useless to point out that the $4 \times 4$
and the $6 \times 6$ systems are inequivalent.
For a given $U$, the $S=1$ phase stability extends to very large positive
values of $V$, while for $V<0$, the triplet-to-singlet transition occurs at
$V_c \simeq -U/2$.
The triangular lattice case is even more remarkable; the triplet phase
stability region is also very extended, with a very steep slope for the
critical line which could be approximated by $V_c \simeq  U/4$ for repulsive $V$. 
But the most surprising feature appears in the attractive $V$ region where a
spectacular re-entrance of the triplet phase for $0.5 \le  U \le 1$ is
observed, showing the complexity of the band structure on frustrated geometries
with the presence of such defect plaquettes.
Once again, even with such a complexity, the UHF appears to be in perfect
quantitative agreement with ED results, hence benchmarking such a mean-field
approach for decorated lattices.
Finally, Fig.~\ref{Fig03} and Fig.~\ref{Fig04} nicely illustrate that
electron-hole symmetry breaking, complex multi-band structure and geometrical
frustration are essential ingredients in order to stabilize and strengthen
magnetic phases \cite{penc,hanisch,arrachea,koretsune}.

\begin{figure}[h] 
\vspace{0.015\textwidth}
\includegraphics[width=0.55\textwidth,clip]{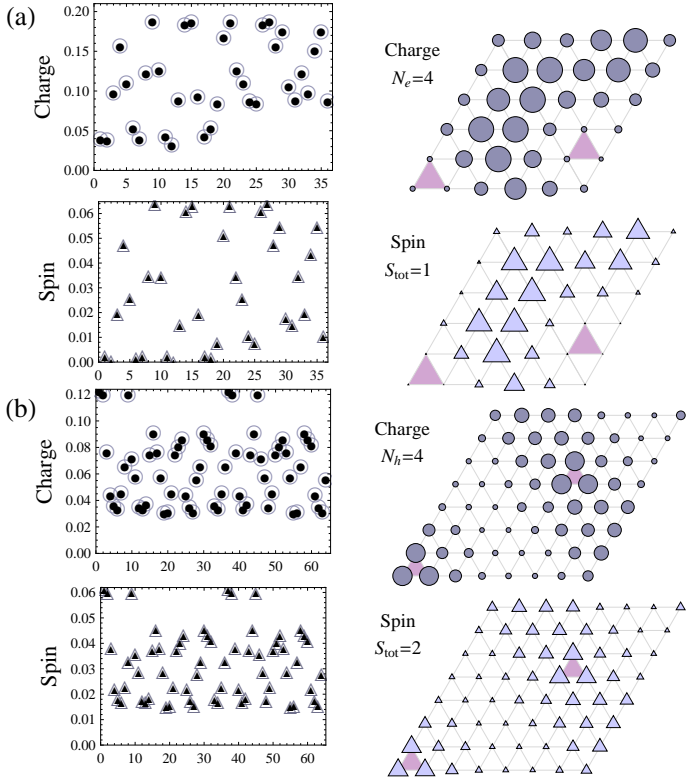}
\caption{
(Color online).
(left column) 
Comparison between ED (small filled symbols) and UHF (large open symbols) results of the local charge (circles) and spin (triangles)
densities as a function of the site index in the case of two defect plaquettes.
(right column) Local charge and spin density snapshots.
(a) $4$  electrons in the 36 site cluster ($S=1$) for $(U,V) = (1.5,1)$.
(b) $4$  holes in the 64 site cluster ($S=2$) for $(U,V) = (1.5,-0.2)$.
\label{Fig05}
}
\end{figure}

\begin{figure}[h] 
\vspace{0.015\textwidth}
\includegraphics[width=0.55\textwidth,clip]{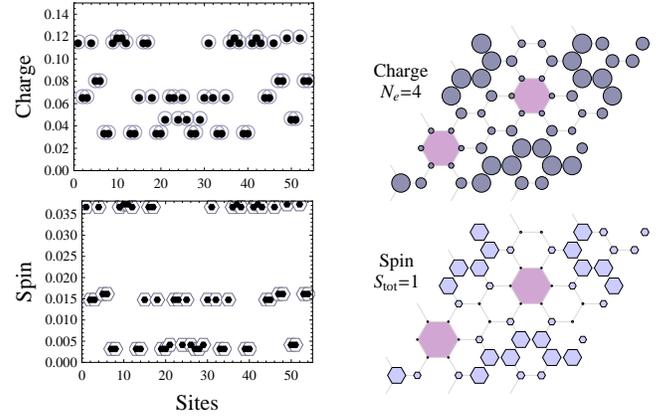}
\caption{
(Color online). Local charge and spin densities in a 54-site hexagonal
lattice as presented in Fig.\ref{Fig05}, for  
$4$  electrons and  $(U,V) = (0.2,0.2)$. The spin ground state is $S=1$.
\label{Fig06}
}
\end{figure}

We now proceed further by focusing our attention on the nature of the GS. For
that purpose, we analyze the local charge and spin densities defined as 
\begin{eqnarray}
\rho_i  &=& \langle n_{i,\uparrow} \rangle + \langle n_{i,\downarrow} \rangle \\ 
s_{i,z} &=& \frac{\langle n_{i,\uparrow} \rangle - \langle n_{i,\downarrow} \rangle}{2}
\end{eqnarray}
for both the triangular and graphene lattices.
The results are depicted in Fig.~\ref{Fig05} and  Fig.~\ref{Fig06}. Prior
to this study, we have computed within the UHF the GS phase diagrams for each
system in order to determine interesting sets of $(U,V)$ parameters;  chosen to
provide a triplet GS in both cases Fig.\ref{Fig05}(a) and Fig.\ref{Fig06} and a quintet in case  Fig.\ref{Fig05}(b).
Note that for the $8 \times 8$ system (b)  we considered only parameters for
which the GS is fully polarized.
Indeed, this spin sector is the only reachable one by ED for such a large system size.
Back to the left panels of Fig.~\ref{Fig05}, where a systematic comparison
between ED and UHF results is done, one can immediately observe a spectacular agreement; a
match of the order of 99.9\% in average has been reached for both local charge
and spin densities. 
The texture of the local charge and spin densities is similar in each case
and exhibits interesting and rich patterns, depending on both the positions of the
defect plaquettes and on the carrier concentration. 
We have considered two different situations in case (a) and (b) of
Fig.~\ref{Fig05}: in (a) the defect plaquette positions break the lattice
(point group and translation) symmetries while some remain preserved in case
(b). We considered electron and hole doped for respectively (a) and (b) and the
same strength for U.  In Fig.~\ref{Fig05}(a), one sees that the local spin
density fluctuates very strongly from site to site and is found very tiny on
the plaquettes sites, although V is not that large.
In Fig.\ref{Fig05}(b) case, one first clearly sees the appearance of stripes
and since the GS is fully polarized, the spin and charge textures are
identical.
Note that this particular stripe structure is a natural consequence of the
symmetry of the plaquette defects on a cluster with periodic boundary
conditions. We have checked that this peculiar stripe structure survives on
larger equivalent clusters (same defect pattern and carrier density). 
In addition one observes a clear dominant moment on the plaquette
sites although the strength of V is only -0.2, which is very small.
Fig.~\ref{Fig06}, on the graphene lattice, is even more remarkable since a
very inhomogeneous and complex GS is obtained even though very small values of
both parameters ($U=V=0.2$) were considered. 
Indeed, for the spin texture, the ratio between the largest and the lowest
local density values (around the defect plaquettes) is at least of the order
$10$. More precisely, the Honeycomb lattice case is the most impressive. The 
fluctuation between the lowest and the highest local densities is at least 
$300\%$.
From the previous two figures, it is now clear that such nano-patterned or
decorated lattices present a very rich and complex physics. Moreover,  the UHF
approach appeared to be spectacularly accurate in the description of the GS
properties which was far beyond our expectations.
It is important to emphasize that we have also checked whether
non-collinear phases could be the GS by including explicitly the
transverse field:
\begin{eqnarray}
\sum_{\sigma} c_{i,\sigma}^+ c_{i,-\sigma} \langle
c_{i,-\sigma}^+ c_{i,\sigma} \rangle - \frac{1}{2} \langle c_{i,\sigma}^+
c_{i,-\sigma} \rangle \langle c_{i,-\sigma}^+ c_{i,\sigma} \rangle.
\end{eqnarray}

In all
investigated cases, only collinear phases have been found.

Let us now show that the accuracy of the UHF is not restricted to weak and
intermediate coupling only.

\begin{figure}[h] 
\vspace{0.015\textwidth}
\includegraphics[width=0.45\textwidth,clip]{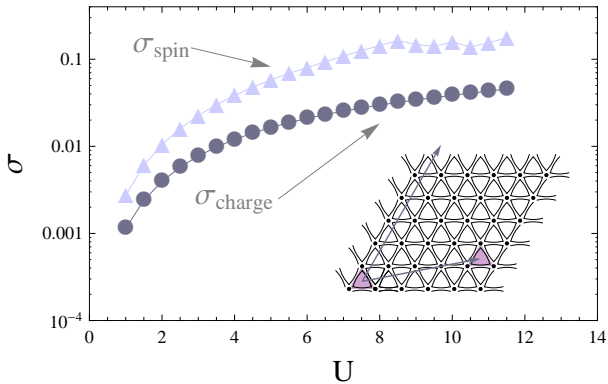}
\caption{(Color online). 
Standard deviation between UHF method and ED for both
the local spin density and local charge as a function of $U$.
The system is a  $6 \times 6$ site cluster on the triangular lattice filled
with 4 electrons. The plaquette potential is set to $V=-1$.
\label{Fig07}
}
\end{figure}

In order to illustrate this, we compare up to large $U$, both local charge and
spin densities obtained within UHF and ED.
For that purpose, we define the following standard deviation $\sigma$:
\begin{eqnarray}
\sigma = \sqrt{ \frac{1}{N_U} \sum_{i} \left( 1 - \Delta_i \right)^2 },
\end{eqnarray} 
where $ \Delta_i = \xi_i^{\textrm{UHF}} / \xi_i^{\textrm{ED}}
$ and $\xi_i$ is either the local spin $s_{i,z}$ or the local
charge $\rho_i$.
We focus on the case of Fig.~\ref{Fig04}(c) ($6 \times 6$ triangular lattice with
two plaquettes and filled with 4 electrons) for $V=-1$ and $U$ ranging from $1$
to $12$.  We recall that for  $U \ge 1$, the total spin is $S= 1$, thus an
average spin per site of $\frac{S}{N} = \frac{1}{36}$.
In our definition, $N_U = N$ for $\xi_i = \rho_i$ whilst for the local spin we
keep only the sites for which $s_{i,z} > \frac{1}{4} \frac{S}{N} = \frac{1}{144}$.
This restriction is introduced in order to avoid insignificantly large numbers
of $\Delta_i$ coming from sites with tiny values of $s_{i,z}$. In other words,
negligible values of $s_{i,z}$ are disregarded. However, typically, $N_U$ is
always at least of the order of $\sim 0.8 N$.
The results are depicted in Fig.\ref{Fig07}. On clearly sees that up to $U =
4$, the agreement is of the order of $99\%$ for both local charge and spin
densities.
As $U$ increases, $\sigma$ increases monotonously and becomes flatter for
large $U$. However, even in this strong coupling  regime ($U=12$),  we obtain an excellent
agreement of the order of $95\%$ and $90\%$ for $\rho_i$ and $s_{i,z}$
respectively.

\begin{figure}[h] 
\vspace{0.015\textwidth}
\includegraphics[width=0.45\textwidth,clip]{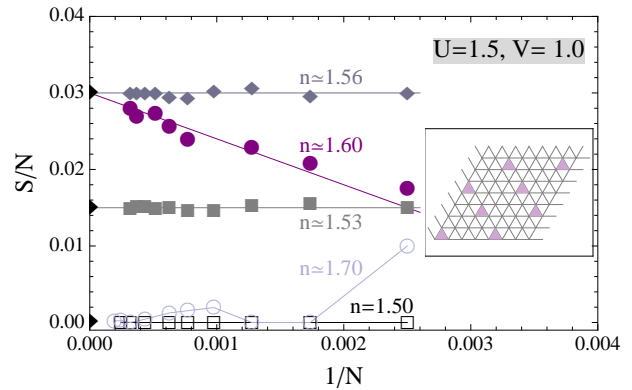}
\caption{(Color online).  Spin ground-state per site $S/N$ as a function of
$1/N$ on a nano-patterned triangular lattice, $N$ being the total number of
sites defined as $N = l\times l$, with $l$ up to $84$. The parameters are $U =
1.5$ and $V = 1$.
The two vectors characterizing the nano-pattern are ${\bf t}_1 = (2,2)$ and
${\bf t}_2 = (-2,2)$ (see Fig.~\ref{Fig01}). The electron concentration $n$
varies from 1.50 to 1.70.
Continuous lines correspond to linear fits of the UHF data (symbols).
\label{Fig08}
}
\end{figure}

One crucial and natural question which arises now is whether the ferromagnetism
survives in the thermodynamic limit. For that purpose, we have computed as
a function of the system size up to $N = 84 \times 84$ sites the magnetic
moment per site $S/N$, for a given pattern on the triangular lattice.  The
results are depicted in Fig.~\ref{Fig08} for various hole concentrations $n_h =
2 - n$ where $1.5 \leq n \leq 1.7$. 
Remark that $n = 1.5$ is particular since the Fermi energy coincides with the
van Hove singularity of the host triangular lattice density of state ($U= V =
0$).
We observe for that particular density a singlet phase $S=0$ for any system
size. 
However, the GS nature changes completely once the hole concentration is
slightly decreased ($n$ increases to $\simeq 1.53$). Indeed, we observe an almost finite
constant $S/N \simeq 0.015$ ratio w.r.t. $1/N$. For example, at $N = 20 \times
20$, a total spin GS of $S=6$ has been found.
As $n$ is further increased to $1.56$, one observes a significant jump of
$S/N\simeq 0.03$ and a similar behavior with the system size, revealing a
strong sensitivity to the hole concentration.
Note that the tendency of the data to weakly deviate from the linear fit has
two possible origins. First, due to commensurability, the carrier density is
set in average, {\it e.g.} for $N = 36 \times 36$, $n= 1.52932$ instead of
$1.53$. Then, as the system size increases and as mentioned before, a very
small change in the hole concentration leads to strong $S/N$ variations.
For $n\simeq1.60$, one now sees a different finite size behavior of $S/N$ which
is  a linear increase as a function of $N$.
It is remarkable to notice that in the thermodynamic limit, $S/N$ coincides with that of  $n \simeq 1.56$.
For the largest hole concentration considered $n=1.70$, we find a surprising
non monotonic behavior of $S/N$. However, in the thermodynamic limit, the spin GS is a singlet.
In the light of these results, the unanticipated strong hole concentration
sensitivity of $S/N$ for such patterns should open interesting perspectives
for spintronic devices.
Indeed, by tuning the carrier density (gate voltage, electron/hole co-doping)
on such 2D nano-patterned lattices, one can switch and manipulate the
magnetic moment of the material.

In conclusion, the competition/interplay between kinetic term, Coulomb
repulsion and correlated on-site disorder leads to very rich and complex
ferromagnetic phase diagrams on nano-patterned / decorated lattices. 
In order to illustrate the richness of the underlying physics, we have
considered systems of various geometries and natures.
Depending on the carrier nature (electron/hole), its concentration, the type of
lattice, the region of stability of ferromagnetism can be very extended, and
exhibits multiple phase transitions with even the possibility of re-entrance in
some cases. The phase diagrams depend strongly on the chosen patterns.
In addition, even in the case of relatively small on-sit potential,  the ground
state appears to be very inhomogeneous. Huge fluctuations of both the local
spin and charge densities have been evidenced.
We have also found that, in the thermodynamic limit, the ground state nature
can change drastically even for a small variation of carrier density. 
The other significant output was to establish that an accessible mean field
type approach, namely the Unrestricted Hartree-Fock method, is a promising and
powerful tool to investigate the many body physics on large scale systems. 
An excellent but unexpected quantitative agreement, even at large $U$, between
UHF and ED have been evidenced.
With the support of both theoretical based approaches (ab-initio/DFT) and the
tremendous expertise in customizing nano-cluster arrays at the atomic precision
\cite{li}, the realization of such nano-patterned compounds is already within
our reach.  Thus, we hope that these findings will open new routes to design
real materials that could be good candidates for spintronic devices.

\acknowledgements
A. R.  acknowledges support by the {\it Agence Nationale de la Recherche} under
grant No. ANR 2010 BLANC 0406-0.


\end{document}